# retinalysis-fundusprep: A python package for robust color fundus image bounds extraction

**Jose Vargas Quiros**[1,2], **Bart Liefers**[1,2], **Karin van Garderen**[1,2], **Eyened Reading Center**[1,2], and **Caroline Klaver**[1,2,3,4]

[1]**Department of Ophthalmology, Erasmus University Medical Center, Rotterdam, the Netherlands**
[2]**Department of Epidemiology, Erasmus University Medical Center, Rotterdam, the Netherlands**
[3]**Department of Ophthalmology, Radboud University Medical Center, Nijmegen, the Netherlands**
[4]**Institute of Molecular and Clinical Ophthalmology, University of Basel, Switzerland**

# Abstract

**Purpose:** Color fundus image (CFI) bounds detection and contrast enhancement are fundamental tasks in automatic CFI analysis. We present an open-source algorithm, published as a Python package, to automatically extract parametric bounds from color fundus images, and for applying contrast enhancement. The software has applications in automated biomarker calculation and AI systems.

**Methods:** Bounds detection was implemented by detecting the CFI's contour via a shortest path algorithm on a polar transformation of the image. A second step detects points along the CFI circle robustly via a circle-fitting RANSAC algorithm. Straight boundaries are detected independently. Finally, the CFI is mirrored along its bounds before contrast enhancement to eliminate edge artifacts. We manually evaluated correctness on the EyeQ and Rotterdam Study datasets.

**Results:** Our open-source Python package allows efficient and accurate bounds detection and contrast enhancement. Our evaluation on the EyeQ CFI quality dataset revealed an error rate of 0.2% for our method, compared to 1.9% for previous work. On a sample of challenging CFI images from the Rotterdam Study, our method did not produce any mistakes.

**Conclusion:** retinalysis-fundusprep is an improvement over existing publicly available code for CFI bounds detection. The improved reliability may allow for its use on a wider range of datasets without the need for manual validation.

**Translational Relevance:** retinalysis-fundusprep enables the use of non-traditional and lower quality CFI images in automated systems. Accuracy of bounds detection directly translates into more accurate biomarkers (eg. vascular density). Its speed enables its use in deep learning training pipelines.

# Introduction

Color fundus images (CFIs) are ubiquitous in ophthalmic research and clinical practice, enabling the visualization and analysis of retinal structures for risk profiling, disease detection, and monitoring. In recent years, deep learning has driven substantial progress in the automated interpretation of CFIs, leading to significant improvements in tasks such as quality assessment, lesion detection, and vessel segmentation[1,2]. However, a critical preprocessing step — accurate and automated detection of the CFI bounds — remains insufficiently addressed by publicly available tools.

Accurate detection of image bounds in CFIs is a critical preprocessing step that directly impacts the reliability of downstream analyses, whether manual, algorithmic, or deep-learning based. The exclusion of irrelevant background regions is essential for the effective application of image enhancement techniques, such as contrast adjustment, histogram equalization, and color balancing. Furthermore, precise bounds delineation facilitates scale estimation and consistent resizing, both of which are fundamental to image standardization. Finally, accurate localization of the bounds enables verification that key anatomical landmarks, such as the optic disc, macula, vascular arcades, or ETDRS regions, are fully contained within the image frame.

While CFI bounds detection most often consists of the detection of a circle, various factors make the problem more difficult in real-world images, namely: non-circular bounds with straight edges, blurry edges, notches and low illumination. For these reasons, designing an algorithm that works robustly and efficiently for a variety of devices and capture conditions is not trivial.

Failure of bounds detection in low quality or non-standard images may result in either their exclusion from automated analysis or the introduction of subtle errors. Even smaller inaccuracies in bounds detection likely have a direct effect on downstream biomarker calculations that rely on the visible retinal area, such as density calculations (e.g. vascular density). Furthermore, systematic inaccuracies may introduce bias in deep learning model training[1].

Previous work has not commonly tackled CFI bounds detection directly. A notable exception is the work by Hashim et al. which presented an algorithm for CFI preprocessing that includes the extraction of a binary mask via Gaussian filtering and global thresholding[3]. The extraction of parametric bounds from the binary mask is, however, not discussed in this work. More recently, pipelines for automated analysis have included CFI preprocessing algorithms. Notably, the vascular biomarker pipeline Automorph[2] includes code for bounds detection through circle detection. The circle is defined by a center and radius and used to crop the image prior to inference. This pipeline has been used on multiple big data analyses on color fundus images[4–6]. Similar code is included in the LWNET vessel segmentation models[7].

In this work, we introduce retinalysis-fundusprep, an open-source Python package for efficient and robust segmentation of CFI boundaries. Our method combines a shortest path algorithm applied to a polar transformation of the image and RANSAC-based circle fitting to accurately delineate the circular fundus region. In addition to circle detection, we implemented detection of straight edges, produced by certain devices. To support downstream analysis, the package includes utilities for cropping to the detected bounds and rescaling images to standardized resolutions. Furthermore, we implement a mirroring-based contrast-enhancement step. This technique synthetically extends the image beyond the detected bounds by reflecting pixel intensities across the boundary. This removes bright halos at the retinal periphery that would otherwise appear in the contrast-enhanced version. Through evaluation on two datasets—the EyeQ image quality dataset[1] and the Rotterdam Study[7]—we demonstrate the accuracy and reliability of our approach, establishing it as a state-of-the-art preprocessing tool for automated CFI analysis.

Our contributions are the following:
- We present an open-source, publicly available Python package for CFI bounds detection. retinalysis-fundusprep is easy to install and use on single or bulk images. It outputs the CFI bounds defined by a circle and optional straight edges if present.
- We present a short evaluation that reveals better performance on (a) the EyeQ dataset[8], a well-known publicly available dataset of CFI images with varying quality levels and (b) Rotterdam Study[7], a large population-based cohort with significant variability in its CFI images, including digitized analog images and several generations of digital devices.

---

[1]

**Methods**

## 0.1 Software development

Our package was developed entirely in Python using standard data science and image manipulation packages: numpy, opencv, Pillow, scikit-learn, skimage. The package has only open-source dependencies.

In this work we present a stable first version of the package. We will follow semantic versioning for future versions.

## 0.2 Input formats

retinalysis-fundusprep operates on image files individually or in bulk. It accepts standard image formats (PNG, JPEG, TIFF) or DICOM format (read with pydicom).

## 0.3 Evaluation

We evaluated the accuracy of the bounds detection via human manual evaluation of the produced masks. For each evaluated algorithm, we overlayed the generated bounds on the image. A grader (JVQ) classified an output as a mistake when the algorithm bounds deviated from the real image bounds in a way that was clearly observable in the overlay.

We compared the accuracy of CFI bounds against the algorithm used in the Automorph vascular analysis software2, which incorporates a CFI preprocessing algorithm capable of extracting circular bounds. Note that for Automorph only mistakes in circle detection were evaluated. In other words, inaccuracies due to the inability of the algorithm to account for straight edges were not counted.

Because retinalysis-fundusprep is a code library, usability tests regarding UI and UX are not applicable.

**Software Summary**

## 0.4 Availability

retinalysis-fundusprep is available at:

https://github.com/Eyened/retinalysis-fundusprep

under an open source license (GNU Affero General Public License v3.0). The package is also available in the Python package index (PyPI) under the name retinalysis-fundusprep.

## 0.5 Usage and outputs

The bounds detection consists of the following steps (see Figure 1):

- **Preprocessing:** The input image is converted to grayscale using the red channel, which typically offers the highest contrast in CFIs. The image is then down-sampled to 256×256 pixels to enable efficient processing.
- **Edge Point Extraction:** The grayscale image is transformed into polar coordinates. Edge features are enhanced by computing the horizontal gradient (via Sobel filtering), followed by attenuation of internal edges using a Gaussian-blurred weighting. This suppresses high-contrast internal structures (e.g., vessels, optic disc) while preserving the peripheral boundary. A shortest-path graph-cut, implemented via dynamic programming, identifies a smooth path through this representation, corresponding to the image boundary.
- **Circular Boundary Fitting:** The extracted edge points are used to fit a circle using a RANSAC-based algorithm. If a sufficient fraction of points (>85%) support the circular model, the boundary is assumed to be circular.
- **Rectangular Edge Detection:** For images with additional rectangular boundaries, RANSAC is used to detect straight horizontal and vertical edges by fitting lines to edge segments within predefined quadrants. A soft constraint ensures that only near-horizontal or near-vertical lines with sufficient support are retained.
- **Final Output:** The resulting geometric parameters—center, radius, and any detected straight edges—are transformed back to the original image scale and used to define the fundus region of interest.

The contrast enhancement consists of the following steps (See Figure 4):

- **Mirroring:** First, horizontal and vertical mirroring is applied along the top, bottom, left, and right edges (if available) by reflecting pixel intensities across the boundary. For the circular border, pixels lying outside the circular region are reflected radially inward based on inverse-normalized radial coordinates.
- **Contrast enhancement:** An unsharp masking procedure is applied: the image is normalized to [0,1] blurred with a Gaussian kernel (with $\sigma$ proportional to the fundus radius), and the high-frequency residual (original minus blurred) is scaled and added back to amplify local contrast. This approach preserves visual consistency near the image boundaries and enhances vascular and structural features within the fundus.

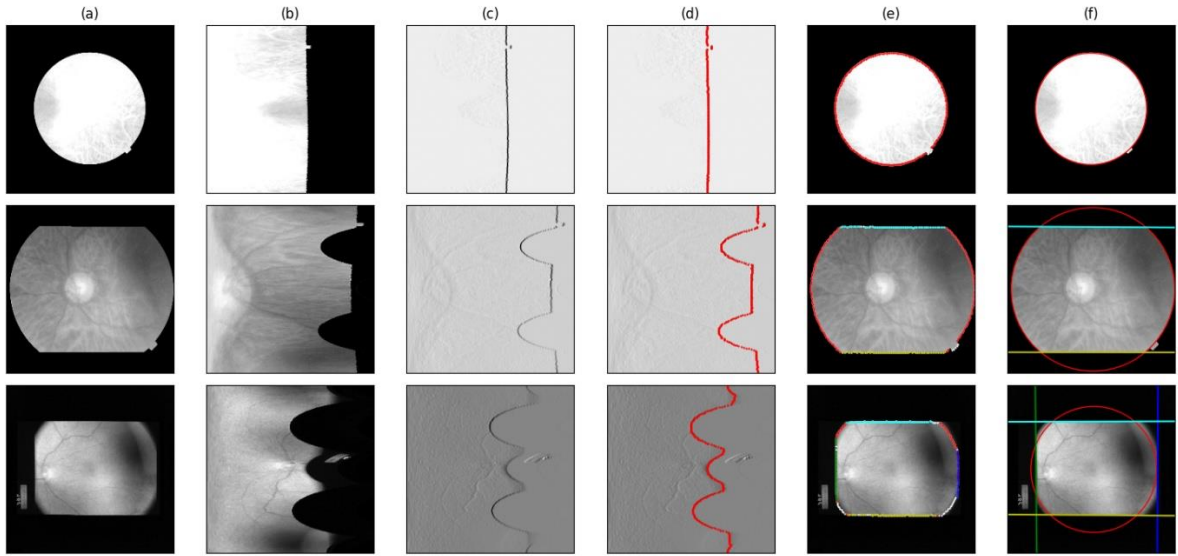

Figure 1. Illustration of the *retinalysis-fundusprep* circle-fitting procedure applied to various input types. (a) Grayscale representation of the input fundus image (red channel). (b) Image transformed into polar coordinates. (c) Horizontal edge features enhanced using the Sobel operator. (d) Output of the shortest path algorithm, identifying boundary candidates. (e) Detected shortest path points projected back into the original Cartesian coordinate space. Red points are utilized by the RANSAC algorithm to robustly fit a circular contour. Cyan, blue, yellow, and green points may optionally be used to fit linear boundaries on the top, right, bottom, and left sides, respectively. (f) Final fitted parametric circle and lines superimposed on the original image. The fitting procedure is robust to notches, artifacts, and other interfering structures.

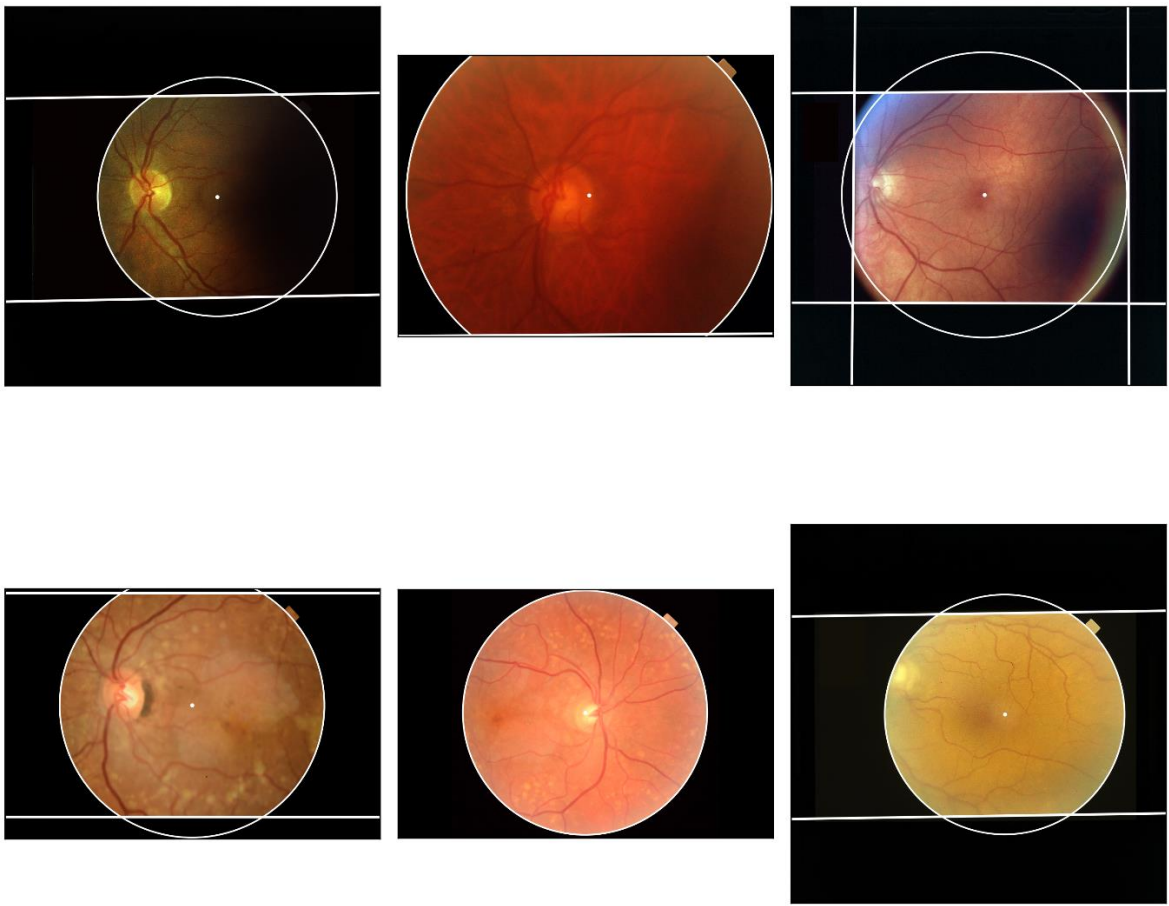

Figure 2. Sample outputs of *retinalysis-fundusprep* on CFIs from the Rotterdam Study cohort.

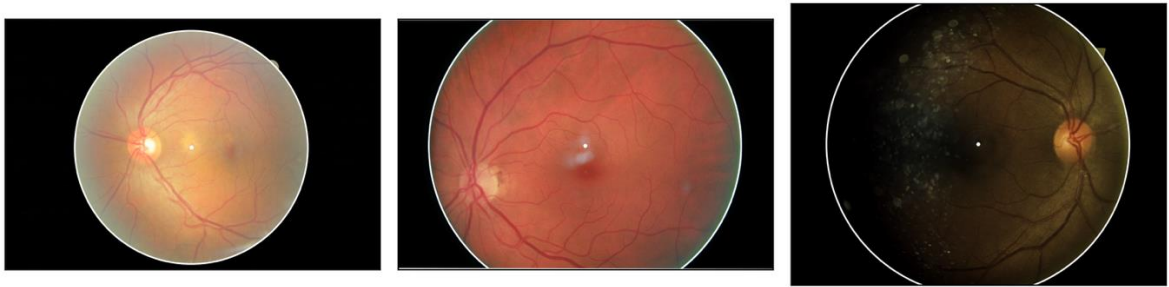

Figure 3. Sample outputs of *retinalysis-fundusprep* on CFIs from the EyeQ dataset.

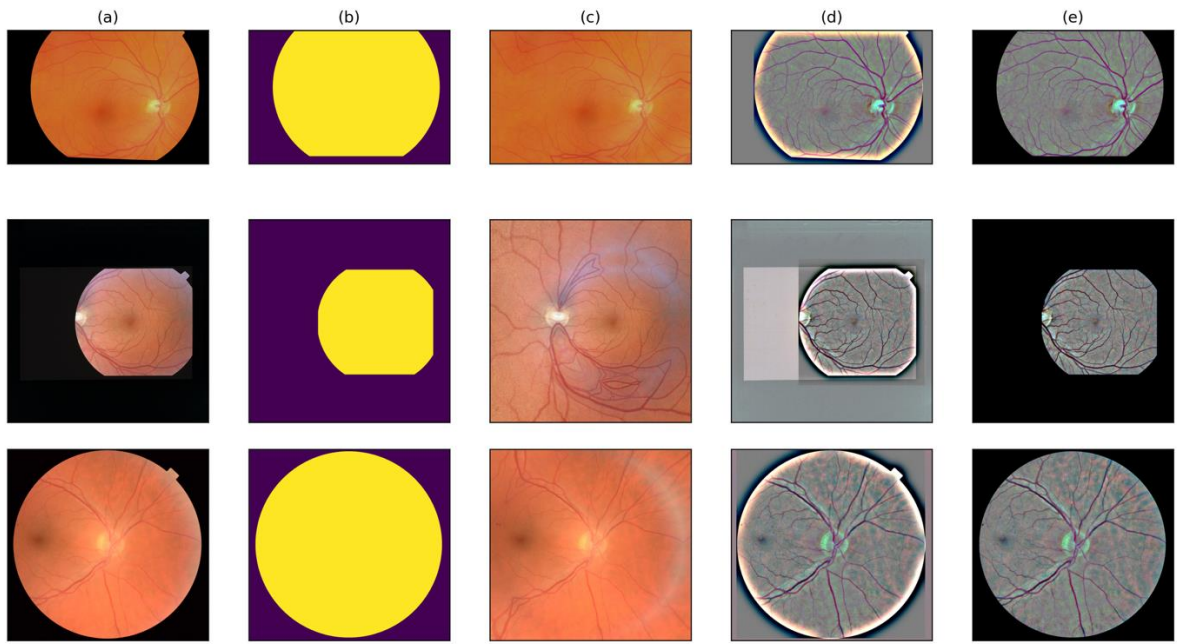

Figure 5. Examples of the contrast enhancement process. (a) Original image. (b) Fundus bounds. (c) The image is mirrored along the fundus bounds. (d) Contrast enhancement without mirroring resulting in a bright halo along the CFI bounds. (e) Contrast-enhanced image, obtained from the mirrored image in (c) without the halo.

**0.6 Evaluation**

Table 1 shows the results of the manual evaluation of mistakes in the output of retinalysis-fundusprep, compared to outputs from the Automorph preprocessing algorithm.

|  | Rotterdam Study | EyePACS EyeQ |
| --- | --- | --- |
| Automorph | 24.1% | 1.9% |
| retinalysis-fundusprep (ours) | 0.0% | 0.2% |

**Table 1.** Results of manual human evaluation of *retinalysis-fundusprep* and baseline's outputs. Percentages indicate error rates, with errors defined as clearly distinguishable mistakes on the fitting of the circle or straight edges.

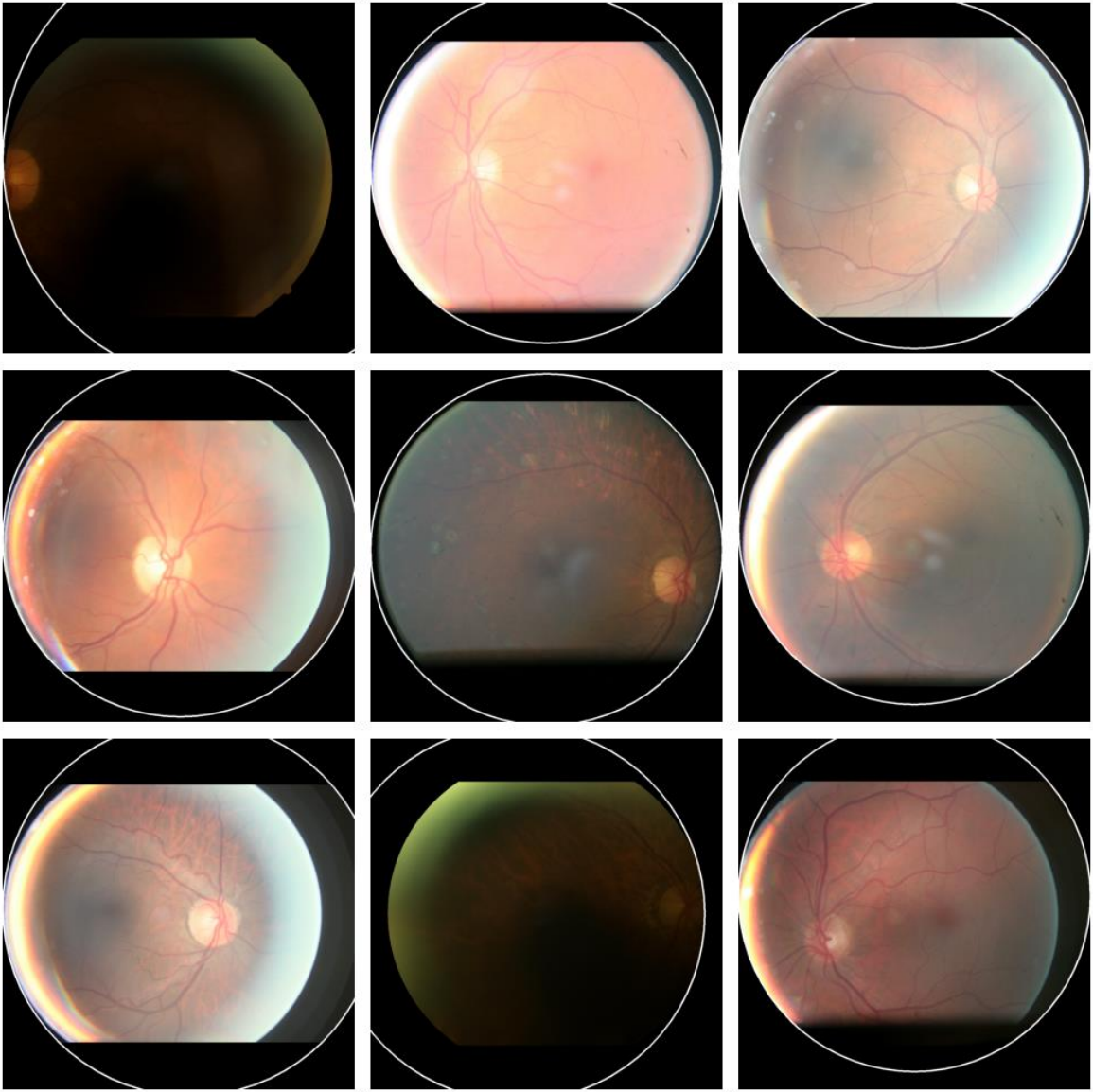

Figure 6. Examples of incorrect bounds found when running Automorph's preprocessing algorithm on EyeQ images.

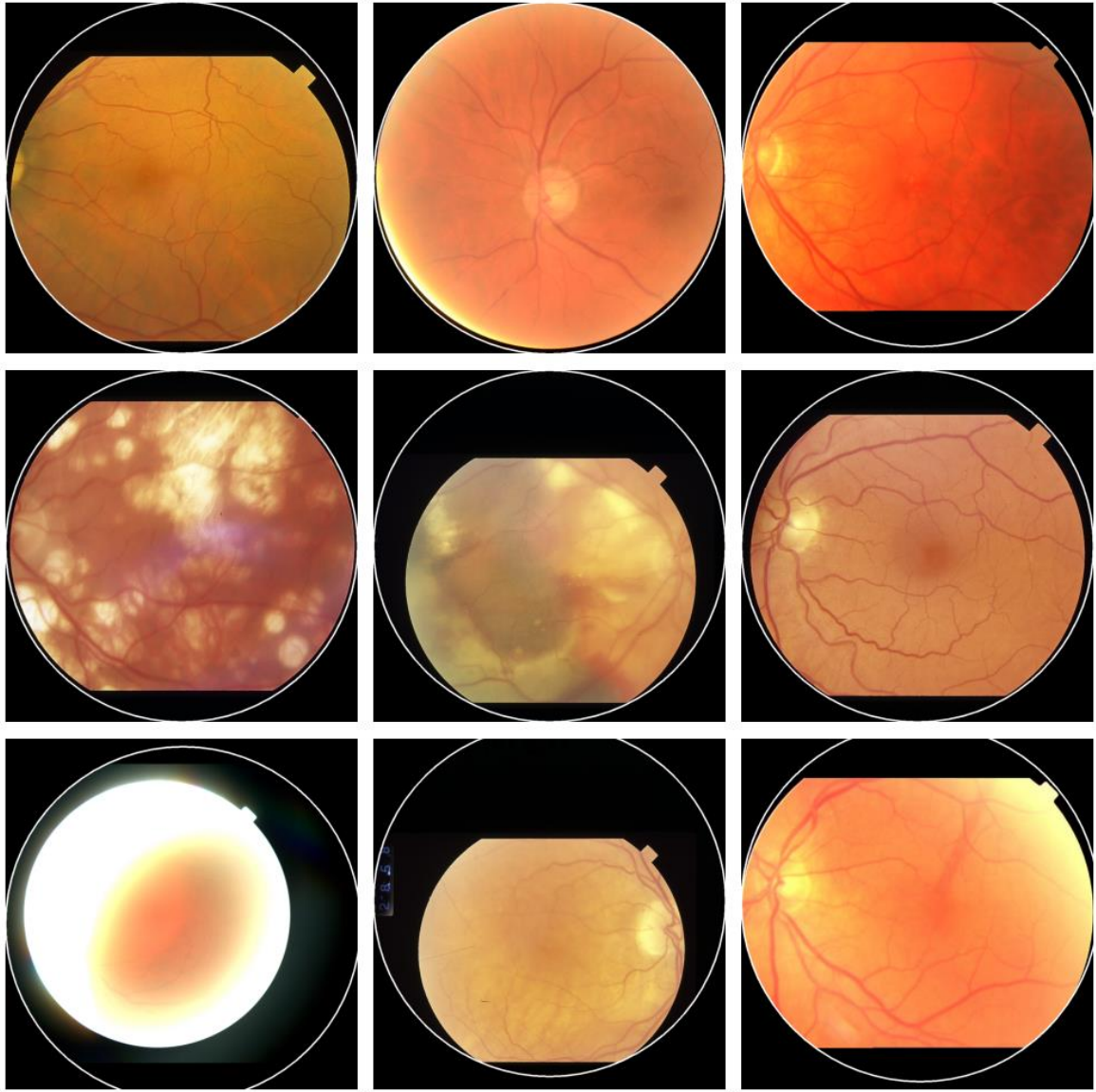

Figure 7. Examples of incorrect bounds found when running Automorph's preprocessing algorithm on Rotterdam Study CFIs.

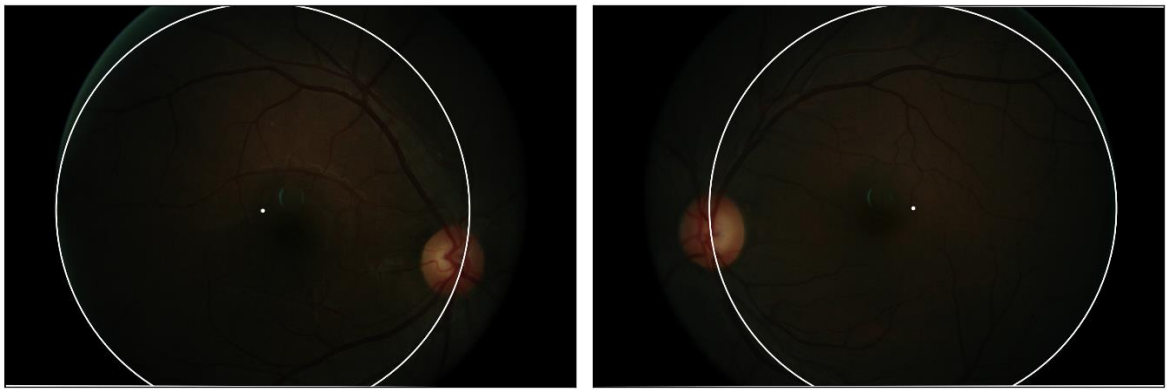

Figure 8. Examples of incorrect bounds found when evaluating *retinalysis-fundusprep* on the EyeQ dataset. The algorithm failed only on very dark images.

## Discussion

Validation of bounds detection outputs on existing datasets revealed issues with previous bounds detection algorithms. The approach evaluated was prone to failure in cases of low illumination, artifacts or straight edges in the CFI mask. Such images are present in real-world CFI datasets and are collected in large cohort studies, as evidenced by our evaluation on the EyeQ dataset and Rotterdam Study images. retinalysis-fundusprep achieves lower and more consistent error rates when compared to these approaches. This is achieved through a novel approach to circular bounds detection via polar transformation of the image.

It must be noted that the very large error rate observed in Rotterdam Study for Automorph is due to the presence of special cases such as bounds with straight edges (Figure 7). Furthermore, the Automorph algorithm tended to fail to ignore small artifacts or slightly illuminated areas in the background of the image. The notch often present on the top right section of CFIs to indicate orientation caused inaccuracies for Automorph but not for retinalysis-fundusprep. retinalysis-fundusprep algorithm was developed to deal with said cases, and it performs nearly perfectly on this set.

The error rates measured in EyeQ are more representative of a diverse, fully digital set with varying image quality levels. Although the difference here is smaller, our algorithm lowers the error rate nearly ten-fold from 1.9% (Automorph) to only 0.2%. This difference may open the door to its application on large datasets with little or no manual validation. Furthermore, the addition of straight edge detection allows the method to operate on a wider range of devices (Figures 2-4).

### 0.7 Potential uses for the software

There are several potential uses for retinalysis-fundusprep:

- As a tool in analysis pipelines, for obtaining preprocessed and enhanced versions of the original images.
- As part of deep learning training and evaluation pipelines, where the images are to be preprocessed and enhanced before being fed into the model.
- As part of other ophthalmic software such as image viewers.

### 0.8 Limitations of the software

While the software provides an improvement over existing approaches in the datasets we evaluated, it still fails for a small fraction (0.2%) of images in the EyeQ dataset. This is important to consider in analyses, particularly those using relatively dark images. Furthermore, although the algorithm was developed and tested on images from various devices and with various artifacts, it may not be able to deal with unseen artifacts or different types of CFI bounds. Application to new datasets may require modifications or calibration.

## Acknowledgments

The authors acknowledge the Sinergia consortium for supporting this project. The authors are grateful to the study participants and the staff from the Rotterdam Study and to all contributors to the (publicly available) datasets that we used in this work. This work was funded by the Swiss National Science Foundation grant no. CRSII5 209510.